\documentclass[11pt,epsf,letterpaper]{article}%
\usepackage{color}
\usepackage{amsmath}
\usepackage{amsfonts}
\usepackage{verbatim}
\usepackage{amssymb}
\usepackage{graphicx}
\usepackage{mathrsfs}%
\providecommand{\U}[1]{\protect\rule{.1in}{.1in}}
\begin{document}

\title{Exact Black Holes and Universality in the Backreaction of non-linear Sigma
Models with a potential in (A)dS4.}
\author{Andr\'{e}s Anabal\'{o}n.\\\textit{Departamento de Ciencias, Facultad de Artes Liberales y}\\\textit{Facultad de Ingenier\'{\i}a y Ciencias, Universidad Adolfo
Ib\'{a}\~{n}ez, Vi\~{n}a del Mar, Chile.}}
\maketitle

\begin{abstract}
The aim of this paper is to construct accelerated, stationary and axisymmetric
exact solutions of the Einstein theory with self interacting scalar fields in
(A)dS4. To warm up, the backreaction of the (non)-minimally coupled scalar
field is solved, the scalar field equations are integrated and all the
potentials compatible with the metric ansatz and Einstein gravity are found.
With these results at hand the non-linear sigma model is tackled. The scalar
field Lagrangian is generic; neither the coupling to the curvature, neither
the metric in the scalar manifold nor the potential, are fixed ab initio. The
unique assumption in the analysis is the metric ansatz: it has the form of the
most general Petrov type D vacuum solution of general relativity; it is a a
cohomogeneity two Weyl rescaling of the Carter metric and therefore it has the
typical Pleba\'{n}ski-Demia\'{n}ski form with two arbitrary functions of one
variable and one arbitrary functions of two variables. It is shown, by an
straightforward manipulation of the field equations, that the metric is
completely integrable without necessity of specifiying anything in the scalar
Lagrangian. This results in that the backreaction of the scalar fields, within
this class of metrics, is universal. The metric functions generically show an
explicit dependence on a dynamical exponent that allows to smoothly connect
this new family of solutions with the actual Plebanski-Demianski spacetime.
The remaining field equations imply that the scalar fields follow geodesics in
the scalar manifold with an affine parameter given by a non-linear function of
the spacetime coordinates and define the on-shell form of the potential plus a
functional equation that it has to satisfy. To further find the exact form of
the potential the simplest case associated to a flat scalar manifold is taken.
The most general potential compatible with the Einstein theory and the metric
anzats is constructed in this case and it is shown that it has less symmetry
than the maximal compact subgroup of the coset construction. Finally, a very
general family of (A)dS4 static hairy black holes is explicitly constructed
and its properties are outlined.

\end{abstract}

\section{Introduction.}

The fact that asymptotically flat black holes are described by a small set of
parameters in four dimensions is what makes them a fundamental object of study
in general relativity and, it is indeed of interest, to see what happens with
this situation when the cosmological constant is included. Actually, to
shortly sumarize this paper, within the class of metrics described below, the
most general backeraction of scalar fields is found and all the possible
potentials compatible with that form of the metric are constructed. It turn
out that this backreaction is described by a small set of parameters, without
any specification of the scalar model and without mentioning the existence of
a black hole. Within the asumptions of this work is proved that the most
general potential for minimally coupled scalar fields compatible with the
Einstein theory is the sum of six exponentials. When the scalar field is
non-minimally coupled, the most general potential can take the form of the sum
of powers of the scalar field. Therefore, an interesting consequence of this
paper is that if some other physical requirement would fix the form of the
potential, the coupling to the curvature will be automatically fixed.

In particular, it is shown that the most general cohomogeneity one, static,
black holes that can be constructed within four dimensional Einstein gravity
with uncharged scalar fields has the following form (in the Einstein frame):%

\begin{equation}
ds^{2}=S(r)(-F(r)dt^{2}+\frac{dr^{2}}{F(r)}+d\Sigma_{k}), \label{M1}%
\end{equation}%
\begin{equation}
S(r)=\frac{\nu^{2}\eta^{\nu-1}r^{\nu-1}}{\left(  r^{\nu}-\eta^{\nu}\right)
^{2}},
\end{equation}

\begin{equation}
F(r)=\left(  \nu^{-1}\left(  1+\frac{2\eta^{\nu}r^{-\nu}}{\nu-2}\right)
r^{2}-\frac{\eta^{2}}{\nu-2}\right)  k+\left(  \left(  \eta^{\nu}-\frac
{r^{\nu}}{\nu+2}+\frac{\eta^{2\nu}r^{-\nu}}{\nu-2}\right)  r^{2}-\frac{\nu
^{2}\eta^{\nu+2}}{\nu^{2}-4}\right)  6M-\frac{\Lambda}{3}. \label{M3}%
\end{equation}
Where $k$ is the constant curvature of the surface $d\Sigma_{k}$, normalized
to be $k=\pm1$ or $0$; $\nu$, $M$, $\Lambda$ and $\eta$ are parameters of the
solution. When $\nu^{2}=1$ this solution is nothing but the Kotler black hole
(also known as the Schwarzschlid-(A)dS black hole). These black holes are
singled out by having a Pleba\'{n}ski-Demia\'{n}ski type completion.

This result holds for any kind of scalar fields and non-linear sigma models
with possitive kinetic energy. The reader can now go to the next section or
read some of the physical motivation behind this study.

\subsection{Motivation.}

It is well known that the introduction of a cosmological constant allow the
black holes to support scalar hair \cite{Martinez:2002ru}. In these cases, the
scalar fields can be considered as a dynamical form of the cosmological
constant, such that when the asymptotic region is reached the scalar field
potential defines the constant value of the curvature. When the cosmological
constant is negative and the asymptotic form of the metric is in the conformal
equivalence class of $R\times S^{2}$, it has been argued that all the
potentials that follow from a class of superpotentials do not give place to
regular hairy black holes \cite{Hertog:2006rr}. In this work it is proved that
is possible to be exhaustive with the potentials within a certain class of
metrics. The interest in being exhaustive is rooted in that is also physically
relevant to clarify what are the possible hairy configurations when the
cosmological constant is positive and when the asymptotic form of the metric
is in any conformal equivalence class when $\Lambda$ is negative.

\subsubsection{$\Lambda>0.$}

The standard cosmological model include a scalar field with a potential
\cite{Naka}. It is therefore a fair question to ask what the stationary and
axisymmetric configurations are, that the backreaction of this scalar field
would generate. This seems to be more than an academic question since the
black hole that is located in the center of our galaxy, Sagittarius A*, is
suffering a close scrutiny \cite{Ghez:2000ay}. In particular, it has been
noted that its angular and quadrupolar momentum, $J$ and $Q$ respectively, can
be determined by the orbital precession of stars very near to it, therefore
allowing to check the relation $Q=-\frac{J^{2}}{Mc}$ that follows from the
Kerr solution \cite{Sadeghian:2011ub}. Due to the uniqueness and ho hair
theorems of asymptotically flat, four dimensional, general relativity this
would actually test whether gravity is described by the Hilbert Lagrangian and
also whether some other fields play an important role in the description of
strong gravitational effects, as has been already studied for the
Einstein-dilaton-Gauss-Bonnet case in \cite{Pani:2009wy}. If the Einstein
equations are guiding the dynamics of the metric in this situation, it could
very well be that the inclusion of a cosmological constant can have a
non-linear effect and a new hairy rotating black hole can be found to describe SgrA*.

To add generality, besides an arbitrary potential, the survey presented here
also let the coupling to the curvature arbitrary. Indeed, it is known that to
encompass a Higgs like potential with an inflationary scenario the Higgs field
should be non-minimally coupled to the curvature \cite{Ferrara:2010yw}.

Static solutions with scalar fields conformally coupled to the curvature have
been explored and there is a family of spherically symmetric and accelerated
black holes. They have a non-trivial potential, whose generalization give rise
to wormholes, regular black holes and bouncing de Sitter cosmologies \cite{A}.
Since all these solutions have been found within assumptions of the form of
the potential and the coupling to the curvature, this work also intend to
explore how generic these solutions are.

Therefore, starting with an arbitrary potential and an arbitrary coupling to
the curvature all the assumptions goes in the symmetries that the problem
should have. Since this work is looking for deviations from Kerr-de Sitter; a
metric that is a cohomogeneity two Weyl rescaling of the Carter metric is
taken as the ansatz. Remarkably, the whole system can be integrated without
any further assumption. All the scalar field potentials compatible with this
metric can be on-shell found. Their off shell construction is explicitly done
in the minimally coupled case while the construction is outlined in the
remaining cases. In the minimally coupled case it is show that the most
general potential is the sum of six exponentials.

With these results at hand the construction of the backreaction of an
arbitrary non-linear sigma model is also done. In this case the scalar fields
follow geodesics in the scalar manifold and the potential satisfy a functional
constraint. In order to solve this constraint, it is necessary to pick a
metric in the scalar manifold and to solve the geodesic motion on it. Since
this is a simple exercise, when doable, it is done just in the minimally
coupled case.

The solutions in general have the Pleba\'{n}ski-Demia\'{n}ski form with a
non-trivial dynamical exponent that (locally) smoothly connect the geometry
with Kerr-de Sitter. This dynamical exponent could be useful to describe the
slow fall off that the galactic rotation curves have at therefore at the end
of the work the spherically symetric black holes are explicitly written. The
same exponent, generically, preclude the interpretation of the metric as a
rotating black hole. However, there is a special case, when the scalar is
conformally coupled, that it does allow for a physically sensible rotating
hairy black hole in (A)dS. This case is briefly discussed here and will be
reported with more details in another publication.

\subsubsection{$\Lambda<0.$}

Dynamical exponents play a central role in the very interesting fact that four
dimensional gravity with a negative cosmological constant and an adequate
matter content is dual to strongly coupled condensed matter phenomena
\cite{Hartnoll:2008vx}, for a recent review see \cite{Hartnoll:2011fn}. The
simplest setting where the situation can be described is for an extremal,
planar, $U(1)$ charged, black hole. In this case the boundary is in the
conformal equivalence class of $R^{3}\,$and it represents the UV behavior of
the theory. The radial coordinate $r$ sets the energy scale. The system is
seen to enjoy a different scale invariance in the UV and the IR. Indeed, at
the boundary, located at $r=0$, the configuration is invariant under the
rescaling $x^{\mu}\rightarrow\lambda x^{\mu}$. On the other hand, since the
near horizon geometry of the spacetime is $AdS_{2}\times R^{2}$, the
configuration in the IR turns out to be invariant under radial and time
rescalings while the remaining coordinates remain unchanged. This can be seen
as the limit when the dynamical exponent goes to infinity of the anisotropic
scalings in space and time that arise in condense matter systems due to the
presence of a Fermi surface \cite{Fermi}. To make the system undergone a
second order phase transition, and to have a finite dynamical exponent, it is
necessary to include charged matter in the gravitational Lagrangian. Either in
the form of spin $0$ or spin $1/2$ particles \cite{Hartnoll:2009sz,
Hartnoll:2010gu}. The class of systems where four dimensional holography is
under control, in the sense of their ultraviolet completion is the ABJM class
\cite{Bagger:2006sk} (for a more recent work see for instance
\cite{Conde:2011sw}) and the Lifshitz solutions of supergravity
\cite{Donos:2010ax}.

The solutions constructed here are independent of the scalar model and they do
have a dynamical exponent, therefore they should correspond as the uncharged
limit of black holes with charged scalars that must exist within the AdS/CM
program. It is interesting to note, that the solutions are asymptotically AdS
in the Fefferman-Graham sense. Thus, it is very likely that its charged
extensions have a non-trivial scaling behavior at the horizon instead of at
infinity as is in the Lifshitz case.

\subsection{Outline.}

The outline of the paper is as follows: in the second section, the metric
ansatz is discussed and its physical relevance recalled. In the third section
the Einstein equations are exactly solved for this ansatz in the presence of a
stationary and axisymetric scalar field with an arbitrary coupling to the
curvature. The most general on-shell potential, compatible with this ansatz,
is found, as well as the most general scalar field that can exist as an exact
solution of this system. Two subsections follow to it; one where the minimally
coupled case is discussed and the most general potential reconstructed and
other when the conformally coupled case is considered and some comments are
made about the existence of a rotating solution. The fourth section is devoted
to non-linear sigma models in general and the same integrability properties
are shown to arise. The problem of finding the exact backreaction of
non-linear sigma models is reduced to the finding of geodesics on the scalar
manifold and a functional equation on the scalar potential. To give a simple
example on how this mechanism work the lineal sigma model is solved. The fifth
section briefly outline the large family of static (A)dS black holes that
arise as exact solutions of this system. The last section discuss the
connection of these solutions with higher dimensional black holes and gauged supergravity.

The notation follows~\cite{wald}. The conventions of curvature tensors are
such that an sphere in an orthonormal frame has positimve Riemann tensor and
scalar curvature. The metric signature is taken to be $(-,+,+,+)$. Greek
letters are in the coordinate tangent space and capital latin letters in the
scalar manifold, $8\pi G=\kappa$ and the units are such that $c=1=\hbar$.

\section{The metric anzats.}

It was Carter \cite{BC} who, requiring the separability of the Klein-Gordon
and Hamilton-Jacobi equations on a background, found the generalization of the
Kerr-Newman metric to the case when the cosmological constant is present,
furthermore adding a NUT parameter to this solution. This metric was
discovered in parallel by Pleba\'{n}ski \cite{Pleba} who wrote it in the Wick
rotated double Kerr-Schild form. The Carter-Pleba\'{n}ski metric, also known
as Kerr-Newman-TAUB-NUT-de Sitter metric contains 7 parameters: the mass, the
NUT parameter, the electric charge, the magnetic charge, the cosmological
constant, the angular momentum and an extra discrete parameter.

The Carter-Pleba\'{n}ski metric has a very important generalization namely the
Pleba\'{n}ski-Demia\'{n}ski spacetime \cite{Plebanski:1976gy}. These
spacetimes contain one further parameter related with a conical singularity
which produce that the two black holes present in the maximal extension of the
spacetime accelerate apart or collapse one against the other, depending
whether there is a conical defect or excess \cite{Bicak:1999sa}. The
non-rotating version of the Pleba\'{n}ski-Demia\'{n}ski metric was discovered,
just a few years after the Schwarzschild solution, by Levi-Civita in 1918, in
the forthcoming years it was rediscovered and analyzed at least three times
(see \cite{Kinnersley:1970zw} for references). Its was baptized after the name
C-metric by Ehlers and Kundt in 1968. Most of its physical life comes from the
interpretation of Kinnersely and Walker \cite{Kinnersley:1970zw} who showed
that it is the general relativistic analogue of the Born solution in
electrodynamics. The election of a good parametrization and analytic extension
of the spacetime allowed them to show that it can be interpreted as two black
holes being accelerated apart. When the cosmological constant vanishes, the
issue of radiation was settled by Bi\v{c}\'{a}k who proved that the Bondi news
are non-trivial for this spacetime, implying that there is a flux of
gravitational radiation through null infinity \cite{flux}.

In five dimensions a Wick rotation of a metric of the
Pleba\'{n}ski-Demia\'{n}ski type has been used as the base space of a
fibration to construct a large family of Ricci flat (and charged) black holes
that contains as special limits the Myers-Perry black hole and also the
Black-Ring \cite{Lu:2008js}. The generalization of the
Pleba\'{n}ski-Demia\'{n}ski spacetime and its static limit, the C-metric, to
higher dimensions is currently unknown, since the obvious generalization fails
(see the discussion in the appendix of \cite{Kubiznak:2008qp}).This class of
metrics have been found to be solution of the four dimensional
Einstein-Maxwell-Conformally coupled scalar field with a quartic potential in
\cite{Anabalon:2009qt}.

The ansatz that allows to obtain the Pleba\'{n}ski-Demia\'{n}ski spacetime in
vacuum general relativity is the starting point of this paper:%

\begin{equation}
ds^{2}=\frac{S(q,p)}{-6+\xi\kappa\phi^{2}}\left(  \frac{1+p^{2}q^{2}}%
{Y(q)}dq^{2}+\frac{1+p^{2}q^{2}}{X(p)}dp^{2}-\frac{Y(q)}{1+p^{2}q^{2}}\left(
p^{2}d\tau+d\sigma\right)  ^{2}+\frac{X(p)}{1+p^{2}q^{2}}\left(  d\tau
-q^{2}d\sigma\right)  ^{2}\right)  ,\label{M}%
\end{equation}
where%

\begin{equation}
\phi=\phi(q,p).
\end{equation}
The choice of the form of the conformal factor in the metric ansatz allows to
integrate the problem when the scalar field is non-minimally coupled.

\section{The single scalar field case.}

To understand how the calculation goes in the most complicated case of the
non-linear sigma model it is useful to do the simpler exercise of the single
scalar field first. It also allows to understand how the non-minimal coupling
to the curvature it is connected with the allowed form of the potentials that
follow from the metric (\ref{M}) and the backreaction of Einstein gravity.

The action principle is:%

\begin{equation}
S(g,\phi)=\int d^{4}x\sqrt{-g}\left[  \frac{R}{2\kappa}-\frac{1}{2}g^{\mu\nu
}\partial_{\mu}\phi\partial_{\nu}\phi-\frac{\xi}{12}\phi^{2}R-V(\phi)\right]
, \label{AP}%
\end{equation}
with field equations:
\begin{equation}
E_{\mu\nu}:=R_{\mu\nu}-\frac{1}{2}g_{\mu\nu}R-\kappa T_{\mu\nu}=0, \label{eqs}%
\end{equation}%
\begin{equation}
T_{\mu\nu}=\partial_{\mu}\phi\partial_{\nu}\phi-\frac{1}{2}g_{\mu\nu}\left(
\partial\phi\right)  ^{2}-g_{\mu\nu}V(\phi)+\frac{\xi}{6}\left(  g_{\mu\nu
}\kern1pt\vbox{\hrule height 0.9pt\hbox{\vrule width 0.9pt\hskip
2.5pt\vbox{\vskip 5.5pt}\hskip 3pt\vrule width 0.3pt}\hrule height
0.3pt}\kern1pt-\nabla_{\mu}\nabla_{\nu}+R_{\mu\nu}-\frac{1}{2}g_{\mu\nu
}R\right)  \phi^{2}%
\end{equation}

\begin{equation}
\kern1pt\vbox{\hrule height 0.9pt\hbox{\vrule width 0.9pt\hskip
2.5pt\vbox{\vskip 5.5pt}\hskip 3pt\vrule width 0.3pt}\hrule height
0.3pt}\kern1pt\phi=\frac{\xi}{6}R\phi+\frac{\partial V}{\partial\phi},
\end{equation}
where $V(\phi)$ is arbitrary. When $\xi=0$ the scalar field is minimally
coupled and when $\xi=1$ it is conformally coupled. The ansatz for the metric
is (\ref{M}) and the scalar field is taken to respect the symmetry of the metric:%

\begin{equation}
\phi=\phi(q,p)
\end{equation}
Let us consider the tensor (\ref{eqs}). The equations $E_{\sigma}^{\tau}$ and
$E_{\tau}^{\sigma}$ are linear in $S(q,p)$ and its partial derivatives. It
turns out that they can be combined to obtain the form of the $\partial_{p}\ln
S$ and $\partial_{q}\ln S$ in terms of of $X$, $Y$, and its derivatives.
Therefore, using the fact that $\left[  \partial_{p},\partial_{q}\right]  \ln
S=0$ it is possible to obtain a relation between $X$, $Y$ and its derivatives.
Since these are functions of independent variables the integration of this
relation is straightforward. The most general solution is:%

\begin{align}
X(p)  &  =C_{0}+C_{2}p^{2}+C_{4}p^{4}+C_{1}p^{-\nu+2}+C_{3}B_{3}p^{\nu
+2},\label{1}\\
Y(q)  &  =C_{4}-C_{2}q^{2}+C_{0}q^{4}+C_{3}C_{1}q^{-\nu+2}+B_{3}q^{\nu+2}.
\label{2}%
\end{align}
Having these metric functions it is possible to integrate $S(q,p)\,$to obtain%

\begin{equation}
S(q,p)=C\frac{p^{\nu-1}q^{\nu-1}}{(C_{3}p^{\nu}+q^{\nu})^{2}}. \label{3}%
\end{equation}
Next, to solve the scalar field we give a look to $E_{\tau}^{\tau}-E_{q}^{q}$
and $E_{\sigma}^{\sigma}-E_{p}^{p}$, and $E_{p}^{q}$ from these equations it
can be extracted the value of $\left(  \partial_{p}\phi\right)  ^{2}$ and
$\left(  \partial_{q}\phi\right)  ^{2}$ and $\left(  \partial_{q}\phi\right)
\left(  \partial_{p}\phi\right)  $. Using (\ref{1})-(\ref{3}) it is possible
to check that:
\begin{align}
E_{\tau}^{\tau}-E_{q}^{q}  &  =0\Leftrightarrow\left(  \partial_{q}%
\phi\right)  ^{2}=\frac{\nu^{2}-1}{12q^{2}\kappa}\frac{\left(  \xi\kappa
\phi^{2}-6\right)  ^{2}}{\left(  \kappa\xi\left(  \xi-1\right)  \phi
^{2}+6\right)  },\label{E1}\\
E_{\sigma}^{\sigma}-E_{p}^{p}  &  =0\Leftrightarrow\left(  \partial_{p}%
\phi\right)  ^{2}=\frac{\nu^{2}-1}{12p^{2}\kappa}\frac{\left(  \xi\kappa
\phi^{2}-6\right)  ^{2}}{\left(  \kappa\xi\left(  \xi-1\right)  \phi
^{2}+6\right)  },\label{E2}\\
E_{p}^{q}  &  =0\Leftrightarrow\partial_{q}\phi\partial_{p}\phi=-\frac{\nu
^{2}-1}{12pq\kappa}\frac{\left(  \xi\kappa\phi^{2}-6\right)  ^{2}}{\left(
\kappa\xi\left(  \xi-1\right)  \phi^{2}+6\right)  }. \label{E3}%
\end{align}
It follows from (\ref{E1})-(\ref{E3}) that:%

\begin{equation}
q\partial_{q}\phi+p\partial_{p}\phi=0\Leftrightarrow\phi=F\left(  \frac{q}%
{p}\right)  \equiv F(z).
\end{equation}
In this case $E_{\tau}^{\tau}-E_{q}^{q},$ $E_{\sigma}^{\sigma}-E_{p}^{p}$ and
$E_{p}^{q}$ reduce to a single equation for the scalar field:%

\begin{equation}
\left(  F^{\prime}\right)  ^{2}=\frac{\left(  \nu^{2}-1\right)  \left(
\xi\kappa F^{2}-6\right)  ^{2}}{12\kappa z^{2}\left(  \kappa\xi\left(
\xi-1\right)  F^{2}+6\right)  } \label{ODE}%
\end{equation}
where $F^{\prime}=\frac{dF}{dz}$. This equation can be exactly integrated and
gives a not very illuminating result for arbitrary $\xi$, of the form:
\begin{equation}
z=z(\phi,\xi,h,\nu) \label{x}%
\end{equation}
where $h$ is an integration constant. At this stage all the diagonal
components of $E_{\nu}^{\mu}$ are equal while the non-diagonal are zero. The
only remaining equation is the one that determines $V$ which is%

\begin{align}
V\left(  \phi,z\right)   &  =\left(  12\kappa C\right)  ^{-1}\,\left[
C_{0}\left(  \nu-1\right)  \left(  \nu-2\right)  z^{\nu+1}+C_{4}\left(
\nu+1\right)  \left(  \nu+2\right)  z^{\nu-1}-4C_{3}\left(  \nu^{2}-1\right)
\left(  C_{0}z+C_{4}z^{-1}\right)  \right. \nonumber\\
&  \left.  +C_{3}^{2}C_{0}\left(  \nu+1\right)  \left(  \nu+2\right)
z^{-\nu+1}+C_{3}^{2}C_{4}\left(  \nu-1\right)  \left(  \nu-2\right)
z^{-\nu-1}\right]  (-6+\xi\kappa\phi^{2})^{2} \label{V}%
\end{align}

Therefore, replacing (\ref{x}) in (\ref{V}) the potential is written uniquely
in terms of $\phi$. It follows that, by explicit construction, all the
potentials and its backreaction on the class of metrics (\ref{M}) have been
determined. Note that when $\nu^{2}=1$ the scalar field is constant, the
potential too and the metric becomes the accelerated version of Kerr-NUT-AdS
found by Pleba\'{n}ski-Demia\'{n}ski \cite{Plebanski:1976gy}, which reduces to
Kerr-(A)dS when $C_{3}=0$ and $C_{1}=0$ $(\nu=1)$. It follows that the class
of metrics studied here are (locally) continuos deformations of Kerr-NUT-AdS.
This is not true globally since the topology is still free at this point.

The static limit and the form of the potentials for the minimally coupled and
conformally coupled cases are described in more detail in the next subsections.

\subsection{The static limit.}

For further reference it is worth to understand the static limit of the metric
(\ref{M}). It is achieved in the scaling limit $x^{\mu}\rightarrow\lambda
x^{\mu}$ and $\lambda\rightarrow0$. Note that the conformal factor transforms
homogenously. Indeed, from (\ref{3}) it follows that $S(\lambda q,\lambda
p)=\lambda^{-2}S(q,p)$. Since the objective is to retain the largest possible
number of integration constants in the limit $\lambda\rightarrow0$, it follows
that the constant in the metric functions (\ref{1}) and (\ref{2}) should
transform as $C_{2}\rightarrow\lambda^{-2}C_{2}$, $C_{1}\rightarrow
\lambda^{\nu-2}C_{1}$ and $B_{3}\rightarrow\lambda^{-2-\nu}B_{3}$. With these
rescalings the $\lambda\rightarrow0$ limit takes the following form:%

\begin{equation}
ds^{2}=\frac{S(q,p)}{-6+\xi\kappa\phi^{2}}\left(  \frac{dq^{2}}{Y(q)}%
+\frac{dp^{2}}{X(p)}-Y(q)d\tau^{2}+X(p)d\sigma^{2}\right)  ,
\end{equation}

\begin{align}
X(p)  &  =C_{0}+C_{2}p^{2}+C_{1}p^{-\nu+2}+C_{3}B_{3}p^{\nu+2},\\
Y(q)  &  =C_{4}-C_{2}q^{2}+C_{3}C_{1}q^{-\nu+2}+B_{3}q^{\nu+2}.
\end{align}
While the scalar field and the potential remains unchanged.

\subsection{Minimally coupled case. ($\xi=0$)}

In this case the differential equation for the scalar field (\ref{ODE}) gives:%

\begin{equation}
\phi=\pm\sqrt{\frac{\nu^{2}-1}{2\kappa}}\ln(h\frac{q}{p}) \label{MCS}%
\end{equation}
where $h$ is a would be hairy integration constant. The potential can be
written in terms of the coefficients $\alpha_{+}=\sqrt{2\kappa}\frac{\nu
+1}{\sqrt{\nu^{2}-1}}=\sqrt{2\kappa}\sqrt{\frac{\nu+1}{\nu-1}}$ and
$\alpha_{-}=\sqrt{2\kappa}\frac{\nu+1}{\sqrt{\nu^{2}-1}}=\sqrt{2\kappa}%
\sqrt{\frac{\nu-1}{\nu+1}}$ as follows:%
\begin{align}
V\left(  \phi\right)   &  =\frac{3}{\kappa C}\,\left[  \frac{C_{0}\left(
\nu-1\right)  \left(  \nu-2\right)  }{h^{\nu+1}}e^{\alpha_{+}\phi}+\frac
{C_{4}\left(  \nu+1\right)  \left(  \nu+2\right)  }{h^{\nu-1}}e^{\alpha
_{-}\phi}-4C_{3}\left(  \nu^{2}-1\right)  \frac{C_{0}}{h}e^{\frac{\alpha
_{+}-\alpha_{-}}{2}\phi}\right. \nonumber\\
&  \left.  +\frac{C_{3}^{2}C_{0}\left(  \nu+1\right)  \left(  \nu+2\right)
}{h^{-\nu+1}}e^{-\alpha_{-}\phi}+\frac{C_{3}^{2}C_{4}\left(  \nu-1\right)
\left(  \nu-2\right)  }{h^{-\nu-1}}e^{-\alpha_{+}\phi}-4C_{3}\left(  \nu
^{2}-1\right)  \frac{C_{4}}{h^{-1}}e^{\frac{\alpha_{-}-\alpha_{+}}{2}\phi
}\right]  . \label{off shell}%
\end{align}

Where the positive branch of (\ref{MCS}) was taken. The negative branch can be
obtained changing $\phi\rightarrow-\phi$ everywhere.

\subsection{Conformally coupled case and a rotating solution. ($\xi=1$)}

In this case the scalar field is:%

\begin{equation}
\phi=\pm\sqrt{\frac{6}{\kappa}}\frac{q^{\mu}-p^{\mu}+h\left(  q^{\mu}+p^{\mu
}\right)  }{q^{\mu}+p^{\mu}+h\left(  q^{\mu}-p^{\mu}\right)  },\qquad\mu
^{2}=\frac{\nu^{2}-1}{3}.
\end{equation}
and the potential is a rational function of $\phi$ obtained replacing%

\begin{equation}
z=\left[  \frac{h-1}{h+1}\frac{\phi\sqrt{\kappa}+6}{\phi\sqrt{\kappa}%
-6}\right]  ^{\frac{1}{\mu}}%
\end{equation}
in (\ref{V}). As before, the negative branch correspond to the transformation
$\phi\rightarrow-\phi$ in the potential. It can be seen that when $\nu^{2}=4$
then $\mu^{2}=1$. In this case ($\nu=2$, $\mu=1$):%

\begin{equation}
\phi=\pm\sqrt{\frac{6}{\kappa}}\frac{q-p+h\left(  q+p\right)  }{q+p+h\left(
q-p\right)  },
\end{equation}
and the denominator of the conformal factor of the metric is:%

\begin{equation}
\frac{S(q,p)}{-6+\xi\kappa\phi^{2}}=\frac{C\left(  p+q-hp+hq\right)  ^{2}%
}{24\left(  h^{2}-1\right)  (C_{3}p^{2}+q^{2})^{2}}%
\end{equation}

Which is regular at $p=0$. Allowing its interpretation as a rotating black
hole \cite{AM2}.

\section{An arbitrary Sigma model.}

The fact that the Einstein equations coupled to scalar fields are exactly
solvable for this class of metrics is a consequence of the decoupling of the
matter sector sector and the gravity sector. When the scalar field is
minimally coupled this is a consequence of the structure of the
energy-momentum tensor of stationary and axisymmetric scalar fields
$T_{\sigma}^{\tau}=0=T_{\tau}^{\sigma}$. When the scalar field is
non-minimally coupled a redefinition of the conformal factor of the metric
ansatz cancels out the contribution from the non-minimal coupling to the
energy momentum tensor. This is a very special property of scalars, for
instance stationary and axisymmetric spin one fields have a energy momentum
tensor that is, in general, such that $T_{\sigma}^{\tau}\neq0\neq T_{\tau
}^{\sigma}$.

It is therefore interesting to extend the complete integrability to a generic
non-linear sigma model to appreciate how a pattern appears. The classical work
where non-linear sigma models where studied, without a potential, in four
dimensions is \cite{Breitenlohner:1987dg}.

The theory under consideration is:%

\begin{equation}
S(g,\phi^{C})=\int d^{4}x\sqrt{-g}\left[  \frac{R}{2\kappa}-\frac{1}{2}%
G_{AB}\left(  \phi\right)  \partial_{\mu}\phi^{A}\partial_{\nu}\phi^{B}%
g^{\mu\nu}-V(\phi)\right]  ,
\end{equation}
where the number of scalars is arbitrary, $G_{AB}\left(  \phi\right)  $
depends in all the $\phi^{C}$ and $V(\phi)$ should, in principle, inherits the
symmetries of $G_{AB}$ but is left arbitrary at this point.

The field equations are:
\begin{equation}
E_{\mu\nu}:=R_{\mu\nu}-\frac{1}{2}g_{\mu\nu}R-\kappa T_{\mu\nu}=0,
\end{equation}%
\begin{equation}
T_{\mu\nu}=G_{AB}\left(  \phi\right)  \partial_{\mu}\phi^{A}\partial_{\nu}%
\phi^{B}-\frac{1}{2}g_{\mu\nu}G_{AB}\left(  \phi\right)  \partial_{\alpha}%
\phi^{A}\partial_{\beta}\phi^{B}g^{\alpha\beta}-g_{\mu\nu}V(\phi)
\end{equation}

\begin{equation}
G_{CB}\kern1pt\vbox{\hrule height 0.9pt\hbox{\vrule width 0.9pt\hskip
2.5pt\vbox{\vskip 5.5pt}\hskip 3pt\vrule width 0.3pt}\hrule height
0.3pt}\kern1pt\phi^{B}=-\frac{1}{2}(\partial_{A}G_{CB}+\partial_{B}%
G_{CA}-\partial_{C}G_{AB})\partial_{\alpha}\phi^{A}\partial_{\beta}\phi
^{B}g^{\alpha\beta}+\partial_{C}V, \label{PLA}%
\end{equation}

Since the scalar fields are minimally coupled, the metric ansatz is slightly different:%

\begin{equation}
ds^{2}=S(q,p)\left(  \frac{1+p^{2}q^{2}}{Y(q)}dq^{2}+\frac{1+p^{2}q^{2}}%
{X(p)}dp^{2}-\frac{Y(q)}{1+p^{2}q^{2}}\left(  p^{2}d\tau+d\sigma\right)
^{2}+\frac{X(p)}{1+p^{2}q^{2}}\left(  d\tau-q^{2}d\sigma\right)  ^{2}\right)
\label{Carter}%
\end{equation}
The scalar fields respect the symmetry of the metric:%

\begin{equation}
\phi^{A}=\phi^{A}(q,p)
\end{equation}
As in the previous section it is possible to completely integrate the metric
functions, therefore obtaining (\ref{1})-(\ref{3}). Using this information,
the following equalities can be extracted from the field equations:%

\begin{align}
E_{\tau}^{\tau}-E_{q}^{q}  &  =0\Leftrightarrow G_{AB}\partial_{q}\phi
^{A}\partial_{q}\phi^{B}=\frac{\nu^{2}-1}{2q^{2}\kappa},\label{4}\\
E_{\sigma}^{\sigma}-E_{p}^{p}  &  =0\Leftrightarrow G_{AB}\partial_{p}\phi
^{A}\partial_{p}\phi^{B}=\frac{\nu^{2}-1}{2p^{2}\kappa},\\
E_{p}^{q}  &  =0\Leftrightarrow G_{AB}\partial_{p}\phi^{A}\partial_{q}\phi
^{B}=-\frac{\nu^{2}-1}{2pq\kappa}. \label{5}%
\end{align}
Using (\ref{4})-(\ref{5}) it follows that:%

\begin{equation}
G_{AB}\left(  q\partial_{q}\phi^{A}+p\partial_{p}\phi^{A}\right)  \left(
q\partial_{q}\phi^{B}+p\partial_{p}\phi^{B}\right)  =0. \label{6}%
\end{equation}
If $G_{AB}$ is Euclidean and invertible (\ref{6}) has a unique solution:%

\begin{equation}
q\partial_{q}\phi^{A}+p\partial_{p}\phi^{A}=0\Leftrightarrow\phi^{A}%
=F^{A}\left(  \frac{q}{p}\right)  \equiv F^{A}(z),
\end{equation}
in which case (\ref{4})-(\ref{5}) become the same equation:%

\begin{equation}
G_{AB}\partial_{z}F^{A}\partial_{z}F^{B}=\frac{\nu^{2}-1}{2z^{2}\kappa}.
\label{CONS}%
\end{equation}
Using this equation the on-shell form of the potential can be found:%

\begin{align}
V\left(  z\right)   &  =-\left(  2\kappa C\right)  ^{-1}\,\left[  C_{0}\left(
\nu-1\right)  \left(  \nu-2\right)  z^{\nu+1}+C_{4}\left(  \nu+1\right)
\left(  \nu+2\right)  z^{\nu-1}-4C_{3}\left(  \nu^{2}-1\right)  \left(
C_{0}z+C_{4}z^{-1}\right)  \right. \nonumber\\
&  \left.  +C_{3}^{2}C_{0}\left(  \nu+1\right)  \left(  \nu+2\right)
z^{-\nu+1}+C_{3}^{2}C_{4}\left(  \nu-1\right)  \left(  \nu-2\right)
z^{-\nu-1}\right]  \label{V2}%
\end{align}
and coincide with the expression (\ref{V}) when $\xi=0$ and rescaling
$C\longrightarrow-6C$. Remarkably, the on-shell form of the potential is
universal, namely, independent of the sigma model.

Replacing the metric functions in the scalar field equations they become:%

\begin{equation}
G_{CB}\left(  \left(  H+J\right)  \partial_{z}^{2}F^{B}+\left(  \frac{H}%
{z}+W\right)  \partial_{z}F^{B}\right)  +\gamma_{CAB}\left(  H+J\right)
\partial_{z}F^{A}\partial_{z}F^{B}-\partial_{C}V=0,
\end{equation}
where $\gamma_{CAB}=\frac{1}{2}(\partial_{A}G_{CB}+\partial_{B}G_{CA}%
-\partial_{C}G_{AB})$ and%

\begin{equation}
J=J(z)=C^{-1}\left(  C_{4}+C_{0}z^{2}\right)  \left(  C_{3}+z^{\nu}\right)
^{2}z^{1-\nu},
\end{equation}

\begin{equation}
W=W(z)=C^{-1}z^{-\nu}\left(  C_{3}z^{\nu}+1\right)  \left(  C_{0}C_{3}\left(
\nu+3\right)  z^{2}+C_{4}C_{3}\left(  \nu-1\right)  -C_{0}\left(
\nu-3\right)  z^{\nu+2}-C_{4}\left(  \nu+1\right)  z^{\nu}\right)  ,
\end{equation}

\begin{equation}
H=\frac{p^{2-\nu}q^{2-\nu}\left(  q^{\nu}+C_{3}p^{\nu}\right)  \left(
B_{3}q^{\nu}p^{\nu}+C_{1}\right)  }{\left(  1+p^{2}q^{2}\right)  p^{4}S(q,p)}.
\label{A}%
\end{equation}

Therefore, the sigma model equation can be divided in a part that depends
uniquely on $z$ and other that, generically, does not. There are two cases
when $H$ does not impose any extra differential condition on the fields. When
$H$ vanishes $(B_{3}=0=C_{1})$ or when $\nu=\pm2$ and $B_{3}=C_{1}$ and $H$
becomes a function of $z$ only. However, these two cases are such that the
Weyl tensor vanishes. In what follows I will focus only in the generic case.

The generic case implies the following field equations:%

\begin{equation}
Z_{C}\equiv G_{CB}\left(  \partial_{z}^{2}F^{B}+\frac{1}{z}\partial_{z}%
F^{B}\right)  +\gamma_{CAB}\partial_{z}F^{A}\partial_{z}F^{B}=0,
\label{geodesic motion}%
\end{equation}

\begin{equation}
T_{C}\equiv G_{CB}\left(  J\partial_{z}^{2}F^{B}+W\partial_{z}F^{B}\right)
+\gamma_{CAB}J\partial_{z}F^{A}\partial_{z}F^{B}-\partial_{C}V=0. \label{T}%
\end{equation}
A cross check to these equations arise from their integrability conditions.
Multiplying both of them by $\partial_{z}F^{C}$ and using (\ref{CONS}) to get
rid of the term $G_{CB}\partial_{z}F^{C}\partial_{z}F^{B}$:%

\begin{equation}
Z_{C}\partial_{z}F^{C}=\frac{d}{dz}\left[  G_{AB}\partial_{z}F^{A}\partial
_{z}F^{B}-\frac{\nu^{2}-1}{2z^{2}\kappa}\right]  , \label{I1}%
\end{equation}

\begin{equation}
T_{C}\partial_{z}F^{C}=\left(  \frac{W}{J}-\frac{1}{z}\right)  \frac{\nu
^{2}-1}{2z^{2}\kappa}-\frac{1}{J}\frac{dV}{dz}. \label{I2}%
\end{equation}
It is possible to verify that (\ref{I1}) is satisfied by the constraint
(\ref{CONS}) and (\ref{I2}) by the potential (\ref{V2}).

The equation (\ref{geodesic motion}) implies that the scalar fields follows
geodesics in the scalar manifold with affine parameter $\omega=\ln(z)$.
Indeed, defining $F^{A}(z)\equiv H^{A}(\omega)$ (\ref{geodesic motion}) becomes:%

\begin{equation}
G_{CB}\left(  \partial_{\omega}^{2}H^{B}\right)  +\gamma_{CAB}\partial
_{\omega}H^{A}\partial_{\omega}H^{B}=0. \label{Affine geodesic motion}%
\end{equation}
Replacing (\ref{Affine geodesic motion}) in (\ref{T}) a linear differential
equation is obtained:%

\begin{equation}
G_{CB}\left(  \left(  \frac{W}{J}-\frac{1}{z}\right)  \partial_{z}%
F^{B}\right)  -J^{-1}\partial_{C}V=0.
\end{equation}
Using (\ref{I2}) it becomes:%
\begin{equation}
G_{CB}\left(  \frac{2z^{2}\kappa}{\left(  \nu^{2}-1\right)  }\frac{dV}%
{dz}\partial_{z}F^{B}\right)  -\partial_{C}V=0. \label{V3}%
\end{equation}

To further extract information from the system is necessary to pick a metric
in the scalar manifold. To gain some insight in the result, the linear sigma
model case is worked out in the next section.

\subsection{$G_{AB}=\delta_{AB}$.}

The geodesic equation (\ref{Affine geodesic motion}) is in this case:%

\begin{equation}
\partial_{\omega}^{2}H^{A}=0\Leftrightarrow F^{A}=c^{A}\ln(z)+\left(
\delta_{CB}c^{C}c^{B}\right)  h^{A}%
\end{equation}
Where the form of the integration constants $h^{A}$ has been chosen for
further transparency. The integration constants, $c^{A}$ satisfy the
constraint (\ref{CONS}):%

\begin{equation}
\delta_{AB}c^{A}c^{B}=\frac{\nu^{2}-1}{2\kappa} \label{cons2}%
\end{equation}
The equation for the potential (\ref{V3}) is:%

\begin{equation}
\left(  \frac{2\kappa}{\left(  \nu^{2}-1\right)  }c^{B}c_{A}-\delta_{A}%
^{B}\right)  \partial_{B}V=0. \label{cons3}%
\end{equation}

Using (\ref{cons2}) it follows that if the scalar manifold has dimension
$D\,\,$then the matrix $\left(  \frac{2\kappa}{\left(  \nu^{2}-1\right)
}c^{B}c_{A}-\delta_{A}^{B}\right)  $ has one null eigenvalue and $D-1$ non
null eigenvalues. The potential is thus constrained to be of the form
\begin{equation}
V=V(c_{A}\phi^{A})=V(c_{A}F^{A}). \label{V4}%
\end{equation}

It can be checked that (\ref{V4}) solves all the constraints (\ref{cons3}).
The explicit form of the potential can now be obtained from%

\begin{equation}
c_{A}\phi^{A}=c_{A}c^{A}\ln(z)+\left(  c_{B}c^{B}\right)  c_{A}h^{A}%
\Leftrightarrow z=\exp\left(  \frac{2\kappa}{\nu^{2}-1}c_{A}\phi^{A}%
-c_{A}h^{A}\right)
\end{equation}
and therefore the off-shell potential has the same structure than
(\ref{off shell}). Note that the combination $c_{A}\phi^{A}$ is invariant
under the transformations that let the vector $c_{A}$ invariant, namely
$SO(D-1)$.

\section{The cohomogeneity one black holes.}

The physical meaning of the solutions just constructed is more transparent if
its cohomogeneity one limit is taken. To obtain it from the metric
(\ref{Carter}) with metric functions (\ref{1})-(\ref{3}) a non symmetric
scaling limit should be taken. It is then possible to obtain the following
family of black holes:

\begin{equation}
ds^{2}=S(r)(-F(r)dt^{2}+\frac{dr^{2}}{F(r)}+d\Sigma_{k}), \label{ADS1}%
\end{equation}%
\begin{equation}
S(r)=\frac{\nu^{2}\eta^{\nu-1}r^{\nu-1}}{\left(  r^{\nu}-\eta^{\nu}\right)
^{2}},
\end{equation}

\begin{equation}
F(r)=\left(  \nu^{-1}\left(  1+\frac{2\eta^{\nu}r^{-\nu}}{\nu-2}\right)
r^{2}-\frac{\eta^{2}}{\nu-2}\right)  k+\left(  \left(  \eta^{\nu}-\frac
{r^{\nu}}{\nu+2}+\frac{\eta^{2\nu}r^{-\nu}}{\nu-2}\right)  r^{2}-\frac{\nu
^{2}\eta^{\nu+2}}{\nu^{2}-4}\right)  6M-\frac{\Lambda}{3}. \label{ADS4}%
\end{equation}
here $k$ is the constant curvature of the surface $d\Sigma_{k}$, normalized to
be $k=\pm1$, $0$. The normalization of \thinspace$r$ is the simplest that I
found to write the solution. It is straightforward to see that the black holes
are asymptotically (A)dS around $r=\eta$.

The on-shell potential has the same structure than in the cohomogenity two case:%

\begin{align}
V_{\nu}(\phi(r))  &  =\frac{\alpha}{2}\left(  \frac{\nu-1}{\nu+2}\eta^{-\nu
-1}r^{1+\nu}+\frac{\nu+1}{\nu-2}\eta^{\nu-1}r^{1-\nu}\right)  -\frac{\alpha
}{2}\left(  \frac{\nu+1}{\nu-2}\eta^{1-\nu}r^{\nu-1}+\frac{\nu-1}{\nu+2}%
\eta^{1+\nu}r^{-1-\nu}\right) \nonumber\\
&  +2\alpha\frac{\nu^{2}-1}{\nu^{2}-4}\left(  \eta^{-1}r-\eta r^{-1}\right)
+\frac{\Lambda\left(  \nu^{2}-4\right)  }{6\kappa\nu^{2}}\left(  \frac{\nu
+1}{\nu-2}\left(  r\eta^{-1}\right)  ^{\left(  \nu-1\right)  }+\frac{\nu
-1}{\nu+2}\left(  r\eta^{-1}\right)  ^{-\left(  1+\nu\right)  }\right)
,\nonumber\\
&  +\frac{2\Lambda}{3\kappa\nu^{2}}\left(  \nu^{2}-1\right)  \left(
r\eta^{-1}\right)  ^{-1}%
\end{align}

\begin{equation}
\alpha=-\frac{\eta^{2}}{\kappa\nu^{2}}\left(  6M\nu^{2}\eta^{\nu}%
+k(\nu+2)\right)  , \label{par}%
\end{equation}

The scalar fields satisfy:
\begin{equation}
G_{AB}\partial_{r}F^{A}\partial_{r}F^{B}=\frac{\nu^{2}-1}{2r^{2}\kappa}%
\end{equation}
plus the field equation:%

\begin{equation}
G_{CB}\kern1pt\vbox{\hrule height 0.9pt\hbox{\vrule width 0.9pt\hskip
2.5pt\vbox{\vskip 5.5pt}\hskip 3pt\vrule width 0.3pt}\hrule height
0.3pt}\kern1ptF^{B}=-\frac{1}{2}(\partial_{A}G_{CB}+\partial_{B}%
G_{CA}-\partial_{C}G_{AB})\partial_{r}F^{A}\partial_{r}F^{B}g^{rr}%
+\partial_{C}V,
\end{equation}
which therefore constraint the potential to have a given form as in the
cohomogeneity two case.

When there is only one minimally coupled scalar field it has the following form:%

\[
\phi=\pm\left(  \frac{\nu^{2}-1}{2\kappa}\right)  ^{1/2}\ln(r)+h
\]
where $h$ is a would be hairy integration constant; it follows that the off
shell form of the potential is the sum of six exponentials. The metric
(\ref{ADS1})-(\ref{ADS4}) is assymptotically (A)dS at $r=1$. Therefore, the
scalar field is everywhere regular as well as the back hole (\ref{ADS1}%
)-(\ref{ADS4}) outside the horizons.

These black holes are very interesting in many regards:

\begin{itemize}
\item They represent the most general black holes that can be constructed in
the presence of stationary and axisymmetric uncharged scalars, having the
Petrov type D completion (\ref{Carter}). These gravitational fields can be
used to describe the departure from the Newtonian behavior.

\item For each value of the parameters there are two different black holes.
When $\Lambda<0$ this observation implies the existence of phase transitions..

\item In this section only the Einstein frame was discussed. Any other frame
correspond to a simple field redefinition of it.

These issues will be discussed in a forthcoming publication.
\end{itemize}

\section{What's next.}

Although the construction of this paper is interesting due to its generality
it is somehow incomplete due to the lack of inclussion of gauge fields in it
and I would like to describe its inclussion and its connection with gauged
supergravity as well as with higher dimensional black holes.

It is very likely that the fact that the problem posed by the integration of
the backreaction of the general class of scalar fields considered here can be
solved due to the hidden symmetries that the metric (\ref{Carter}) has. This
class of metrics are Petrov type D and support a rank two, irreducible
conformal Killing tensor. The requirement that the metric support a
non-trivial rank two conformal Killing tensor seems to not be very
restrictive. On one hand, all the Petrov type D metrics in vacuum were
constructed by Kinnersley \cite{PetrovD}. On the other hand, the most general
metric admiting an irreducible rank two Killing tensor and $D-2$ Killing
vectors was kinematically constructed by Benenti and Francaviglia \cite{BF}.
The restriction of this family of metrics to be of Petrov type D, plus a
further restriction of some functions, indeed coincide with the form of the
Carter-Pleba\'{n}ski metric in four dimensions. Therefore a cohomegeneity two
Weyl rescaling of it has the Pleba\'{n}ski-Demia\'{n}ski form, cf.
(\ref{Carter}). When the Ricci flat condition is imposed on it gives rise to
all the metrics classified by Kinnersley as special limits \cite{GP}. It
follows that all the Petrov type D, Ricci flat, stationary and axisymmetic
spacetimes admit a (possibly conformal) Killing tensor.

These are, however, not the most general Petrov type D metrics that can be
constructed in four dimensions in presence of matter. It also exist other
solutions when $U(1)$ vector fields are considered \cite{Chong:2004na}. These
solutions, however, do not reduce to the class studied here in the limit of
vanishing vectors and therefore it should exist a family of solutions that
encompass them all. It is worth mentioning that very recently, an extremely
interesting approach has been introduced in the literature of exact solutions
\cite{Houri:2012eq}. It allows to construct general classes of exact solutions
to the Einstein equations with scalars and vector fields in (A)dS, in such a
way that the Hamilton-Jacobi equation is separable for a charged particle. It
can be that the canonical form of the metric thus constructed can be
generalized to admit solutions of the Pleba\'{n}ski-Demia\'{n}ski form by Weyl
rescaling it.

Besides the discussions already mentioned in the body of this work, the study
of metrics admiting a conformal Killing tensor can be of utmost interest for
the construction of higher dimensional black holes with non-trivial
topologies. Remarkably enough, the rather non-trivial metric of the double
spinning black ring is such that the four dimensional base of its Kaluza-Klein
fibration has a conformal Killing tensor. This base space is not of Petrov
type D as follows from the fact that it does not admit a conformal
Killing-Yano tensor unless the single spinning limit is taken
\cite{Durkee:2008an}. Indeed, starting from an analytic continuation of the
Ricci flat solution of \cite{Lu:2008js}, is possible to obtain the following
four dimensional solution%

\begin{equation}
ds^{2}=\frac{a_{0}qp}{\left(  1-\alpha q^{2}p^{2}\right)  ^{2}}\left[
\frac{q^{2}+p^{2}}{Y(q)}dq^{2}+\frac{q^{2}+p^{2}}{X(p)}dp^{2}-\frac
{Y(q)\left(  d\tau-p^{2}d\sigma\right)  ^{2}}{q^{2}+p^{2}}+\frac{X(p)\left(
d\tau+q^{2}d\sigma\right)  ^{2}}{q^{2}+p^{2}}\right]  , \label{C1}%
\end{equation}

\begin{equation}
X(p)=-a_{0}^{2}\alpha^{2}p^{6}+a_{4}\alpha p^{4}+a_{2}p^{2}+a_{3}-a_{0}%
^{2}p^{-2},\qquad Y(q)=a_{0}^{2}\alpha^{2}q^{6}-a_{3}\alpha q^{4}-a_{2}%
q^{2}-a_{4}+a_{0}^{2}q^{-2},
\end{equation}

\begin{equation}
A=\frac{d\tau}{q^{2}p^{2}}+\frac{q^{2}-p^{2}}{q^{2}p^{2}}d\sigma,\qquad
\Phi=\ln(a_{0}qp). \label{C3}%
\end{equation}

Where the normalization of the fields is given by the action principle%

\begin{equation}
S(g,A,\Phi)=\int d^{4}x\sqrt{-g}\left[  R-\frac{1}{4}e^{3\Phi}F^{2}-\frac
{3}{2}\left(  \partial\Phi\right)  ^{2}\right]  , \label{EMD}%
\end{equation}

with field equations%

\begin{align}
R_{\mu\nu}-\frac{1}{2}g_{\mu\nu}R  &  =\frac{3}{2}\left(  \partial_{\mu}%
\Phi\partial_{\nu}\Phi-\frac{1}{2}g_{\mu\nu}\left(  \partial\Phi\right)
^{2}\right)  +\frac{e^{3\Phi}}{2}\left(  F_{\mu\alpha}F_{\nu}^{\cdot\alpha
}-\frac{1}{4}g_{\mu\nu}F^{2}\right)  ,\\
\partial_{\mu}\left(  \sqrt{-g}e^{3\Phi}F^{\mu\nu}\right)   &  =0,\qquad
\frac{1}{\sqrt{-g}}\partial_{\mu}\left(  \sqrt{-g}g^{\mu\nu}\partial_{\nu}%
\Phi\right)  -\frac{1}{4}e^{3\Phi}F^{2}=0.
\end{align}
the non-standard normalization of the scalar field in (\ref{EMD}) is to avoid
the appearance of $\sqrt{3}$ factors in the solution of the field equations.
When the solution (\ref{C1}-\ref{C3}) is oxidized back to five dimensions the
metric contains, in different limits, the Myers-Perry black hole, the
Emparan-Reall black ring, the static and stationary black lens, the
Kaluza-Klein monople and the Schwarzschild-Tangherlini black hole.

When the configurations are compactified, they can be extended to be a
solution of the four dimensional $U(1)^{4}$ gauged supergravity, which is a
truncation of the $\mathcal{N}=8,$ $SO(8)$ gauged supergravity; it is
$\mathcal{N}=2$ supergravity coupled to $3$ abelian vector multiplets. When
the axions are set to zero and only one of the $U(1)$ gauge fields is retained
a further consistent truncation can be obtained%

\begin{equation}
S(g,A,\Phi)=\int d^{4}x\sqrt{-g}\left[  R-\frac{1}{4}e^{3\Phi}F^{2}-\frac
{3}{2}\left(  \partial\Phi\right)  ^{2}+\frac{6}{l^{2}}\cosh(\Phi)\right]  ,
\label{S1}%
\end{equation}
with field equations%

\begin{align}
R_{\mu\nu}-\frac{1}{2}g_{\mu\nu}R  &  =\frac{3}{2}\left(  \partial_{\mu}%
\Phi\partial_{\nu}\Phi-\frac{1}{2}g_{\mu\nu}\left(  \left(  \partial
\Phi\right)  ^{2}-\frac{4}{l^{2}}\cosh(\Phi)\right)  \right)  +\frac{e^{3\Phi
}}{2}\left(  F_{\mu\alpha}F_{\nu}^{\cdot\alpha}-\frac{1}{4}g_{\mu\nu}%
F^{2}\right)  ,\\
\partial_{\mu}\left(  \sqrt{-g}e^{3\Phi}F^{\mu\nu}\right)   &  =0,\qquad
\frac{1}{\sqrt{-g}}\partial_{\mu}\left(  \sqrt{-g}g^{\mu\nu}\partial_{\nu}%
\Phi\right)  +\frac{2}{l^{2}}\sinh(\Phi)-\frac{1}{4}e^{3\Phi}F^{2}=0.
\label{S3}%
\end{align}
It is clear that the theory \eqref{S1} coincides with the
Einstein-Maxwell-dilaton theory (\ref{EMD}) when the cosmological constant
vanishes. A trick in the construction of solutions in gauged supergravity
arise from the observation that for metrics with the Carter form the field
equations for the $U(1)$ gauge fields are actually independent of the metric
structure functions ($X,Y$), therefore the ansatz to extend a solution from
the ungauged to the gauged case is to retain the matter fields and integrate
the metric structure functions with the expectation that a solution in the
gauged case can be obtained. This technique has been proven to be very
successful in a large number of cases \cite{Chong:2004na}.

Indeed, the theory \eqref{S1} admit a solution of the form:%

\begin{equation}
ds^{2}=\frac{a_{0}qp}{\left(  1-a_{0}^{2}q^{2}p^{2}\right)  ^{2}}\left[
\frac{q^{2}+p^{2}}{Y(q)}dq^{2}+\frac{q^{2}+p^{2}}{X(p)}dp^{2}-\frac
{Y(q)\left(  d\tau-p^{2}d\sigma\right)  ^{2}}{q^{2}+p^{2}}+\frac{X(p)\left(
d\tau+q^{2}d\sigma\right)  ^{2}}{q^{2}+p^{2}}\right]  , \label{G1}%
\end{equation}

\begin{equation}
X(p)=-a_{0}^{6}p^{6}+a_{4}a_{0}^{2}p^{4}+a_{2}p^{2}+a_{3}-a_{0}^{2}%
p^{-2},\qquad Y(q)=a_{0}^{6}q^{6}-a_{3}a_{0}^{2}q^{4}-a_{2}q^{2}-a_{4}%
+a_{0}^{2}q^{-2}+\frac{1}{4l^{2}a_{0}^{2}}\left(  1+a_{0}^{2}q^{4}\right)  ,
\end{equation}

\begin{equation}
A=\frac{d\tau}{q^{2}p^{2}}+\frac{q^{2}-p^{2}}{q^{2}p^{2}}d\sigma,\qquad
\Phi=\ln(a_{0}qp). \label{G3}%
\end{equation}

Which coincides with the configuration (\ref{C1}-\ref{C3}) when $\frac
{1}{l^{2}}=0$ and $\alpha=a_{0}^{2}$. This provides an embedding of all the
five dimensional black Ricci flat black holes in gauged supergravity and
coincide with the class of metrics studied in this paper when the gauge field
vanishes and a the dynamical exponent, $\nu$ is set to $\nu=\pm2$. In a
forthcoming work the most general solution with gauge fields and arbitrary
$\nu\,\ $will be reported.

\section{Acknowledgments.}

The author would like to thank Guillaume Bossard, Roberto Emparan, Marc
Henneaux, Hideki Maeda, Alfonso Ramallo, Harvey Reall and Simon Ross and for
enlightning discussions and correspondence. A.A. wishes to thank Myriam
Gistelinck for her careful reading of the manuscript. Research of A.A. is
supported in part by the Alexander von Humboldt foundation and by the Conicyt
grant Anillo ACT-91: \textquotedblleft Southern Theoretical Physics
Laboratory\textquotedblright\ (STPLab).


\end{document}